\documentclass[prd,nofootinbib,preprint,showpacs,showkeys,superscriptaddress,doublespace]{revtex4}
\usepackage[english]{babel}
\usepackage{amsmath,amssymb}
\usepackage[dvips]{graphicx}
\usepackage{ifthen,array}
\usepackage[sort&compress]{natbib}
\usepackage{amsfonts}
\usepackage{bbm}
\usepackage{enumerate}

\def\e6{$E(6)$}
\def\10{$SO(10)$}
\def\21{SU(2) $\otimes$ U(1) }

\def\422{$SU(4) \otimes SU(2) \otimes SU(2)$}
\def\321{SU(3) $\otimes$ SU(2) $\otimes$ U(1)}

\newcommand{\nn}{\nonumber}

\def\vev#1{\left\langle #1\right\rangle}

\def\mathbf#1{\hbox{\bf #1}}
\def\textrm#1{\hbox{#1}}

\def\lsim{\raise0.3ex\hbox{$\;<$\kern-0.75em\raise-1.1ex\hbox{$\sim\;$}}}
\def\gsim{\raise0.3ex\hbox{$\;>$\kern-0.75em\raise-1.1ex\hbox{$\sim\;$}}}
\newcommand {\ignore}[1]{}

\newcommand{\AddrAHEP}{%
  AHEP Group, Instituto de F\'{\i}sica Corpuscular --
  C.S.I.C./Universitat de Val{\`e}ncia \\
  Edificio Institutos de Paterna, Apt 22085, E--46071 Valencia, Spain}
\newcommand{\AddrWU}{%
  Deutsches Elektronen-Synchrotron DESY,
  Hamburg, Germany}
\newcommand{\AddrIoa}{
Division of Theoretical Physics, University of Ioannina GR-45110
Ioannina, Greece}

\def\lsim{\mbox{${}^< \hspace*{-7pt} _\sim$}}
\def\gsim{\mbox{${}^> \hspace*{-7pt} _\sim$}}

\def\lfv{$L_f\hspace{-0.95em}/\hspace{0.5em}$ }

\def\m{$\mu^--e^-$ }
\begin{document}
\preprint{IFIC/05-66}

\vspace*{2cm} \title{Enhanced $\mu^--e^-$ conversion in nuclei in the
  inverse seesaw model}

\author{F. Deppisch} \email{frank.deppisch@desy.de}\affiliation{\AddrWU}
\author{T.S. Kosmas} \email{valle@ific.uv.es}\affiliation{\AddrIoa}
\author{J.~W.~F.~Valle} \email{valle@ific.uv.es}\affiliation{\AddrAHEP}

\vspace*{1.0cm}

\begin{abstract}
  We investigate nuclear $\mu^--e^-$ conversion in the framework of an
  effective Lagrangian arising from the inverse seesaw model of
  neutrino masses.  We consider lepton flavour violation interactions
  that arise from short range (non-photonic) as well as long range
  (photonic) contributions. Upper bounds for the \lfv -
  parameters characterizing \m conversion are derived in the inverse seesaw model
  Lagrangian using the available limits on the $\mu^--e^-$ conversion
  branching ratio, as well as the expected sensitivities of upcoming
  experiments. We comment on the relative importance of these two
  types of contributions and their relationship with the measured
  solar neutrino mixing angle $\theta_{12}$ and the dependence on
  $\theta_{13}$. Finally we show how the \lfv $\mu^--e^-$ conversion
  and the $\mu^-\to e^-\gamma$ rates are strongly correlated in this
  model.
\end{abstract}


\maketitle

\section{Introduction}

\vspace*{2mm} The discovery of neutrino oscillations
\cite{fukuda:1998mi,ahmad:2002jz,eguchi:2002dm} shows that
neutrinos are massive~\cite{Maltoni:2004ei} and that lepton
flavour is violated in neutrino propagation. The violation of this
conservation law could show up in other contexts, such as rare
lepton flavour violating (LFV) decays of muons and taus, e.g.
$\mu^-\to e^-\gamma$.  In fact, there are strong indications from
theory that this may be the case. Among the lepton flavour
violating (\lfv) processes, the electron- and muon-flavour
violating nuclear conversion
\begin{equation}
\mu^- + (A,Z) \longrightarrow  e^- \,+\,(A,Z)^*\, ,
\label{I.1}
\end{equation}
is known to provide a very sensitive probe of lepton flavour
violation
\cite{Hisano:1995cp,Kosmas:1993ch,Kosmas:1990tc,Kosmas:2001mv,Kosmas:2001ia,faessler:2000pn,Kitano:2003wn}.
This follows from the distinct feature of a coherent enhancement
in nuclear \m conversion.  From the experimental viewpoint,
currently the best upper bound on the \m conversion branching
ratio comes from the SINDRUM II experiment at PSI
\cite{vanderSchaaf:2003ti}, using $^{197}$Au as stopping target,
\begin{equation}
\label{eq:SINDRUM}
R_{\mu e}^{Au} \leq 5.0\times 10^{-13}\ \ \ \ \rm{90\% C.L.} \ \ \
\end{equation}

The proposed aim of the MECO experiment, the \m conversion
experiment at Brookhaven \cite{Molzon:1998kg}, with $^{27}$Al as
target is expected to reach~\cite{Molzon:1998kg}
\begin{equation}
\label{eq:MECO}
R_{\mu e}^{Al} \leq 2\times 10^{-17}\ \ \ \ \ \ \ \
\end{equation}
about three to four orders of magnitude better than the present best
limit.

An even better sensitivity is expected at the new \m conversion
PRISM experiment at Tokyo, with $^{48}$Ti as stopping target. This
experiment aims at~\cite{Kuno:2000kd}
\begin{equation}
  \label{eq:PRISM}
R_{\mu e}^{Ti} \leq 10^{-18}.
\end{equation}
Such an impressive sensitivity can place severe constraints on the
underlying parameters of \m conversion.

\begin{figure}[t]
\centering
\includegraphics[clip,width=0.9\linewidth]{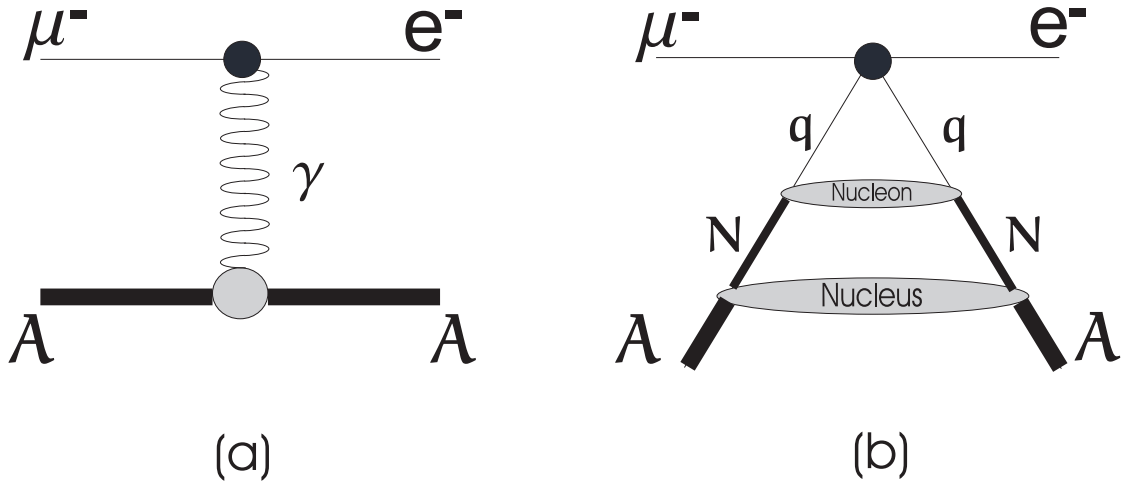}
\caption{ (a) Photonic (long-distance) and (b) non-photonic
  (short-distance) contributions to the nuclear $\mu^-- e^-$
  conversion.}
     \label{fig:Diagrams}
\end{figure}

There are many mechanisms beyond the Standard Model that could
lead to lepton flavour violation (see
\cite{Kosmas:1993ch,Hisano:1995cp,Kosmas:1990tc,Kosmas:2001ia,faessler:2000pn}
and references therein). The corresponding Feynman diagrams can be
classified according to their short-range or long-range character
into two types: photonic and non-photonic, as shown in Fig.
\ref{fig:Diagrams}.
The long-distance photonic mechanisms in Fig.
\ref{fig:Diagrams}(a) are mediated by virtual photon exchange
between nucleus and the $\mu-e$ lepton current. The hadronic
vertex is characterized in this case by ordinary electromagnetic
nuclear form factors. Contributions to $\mu-e$ conversion arising
from virtual photon exchange are generically correlated to $\mu\to
e \gamma$ decay.

The short-distance non-photonic mechanisms in Fig.
\ref{fig:Diagrams}(b) include effective 4-fermion quark-lepton
\lfv interactions which couple the quarks and leptons via heavy
intermediate particles ($W,Z$, Higgs bosons, supersymmetric
particles, etc.) at the tree level, at the 1-loop level or via box
diagrams. The various mechanisms can significantly differ in many
respects, in particular, in what concerns nucleon and nuclear
structure treatment. As a result, they must be treated on a
case-by-case basis.

In this paper we consider \m conversion in the context of a variant of
the seesaw model~\cite{Minkowski:1977sc}, called inverse
seesaw~\cite{mohapatra:1986bd}.  It differs from the standard one in
that no large mass scale is necessary, providing a simple framework
for enhanced \lfv rates, unsuppressed by small neutrino
masses~\cite{bernabeu:1987gr,branco:1989bn}.  The enhancement of \lfv
rates holds in this model even in the absence of supersymmetry and in
the absence of neutrino masses.  For this reason it plays a special
role. For simplicity here we neglect possible supersymmetric
contributions to the \lfv rates that could exist in this model, see
\cite{Deppisch:2004fa}.
Other seesaw constructions with extended gauge groups have been
considered recently, using either left--right gauge
symmetry~\cite{Akhmedov:1995vm} or full SO(10)
unification~\cite{Barr:2005ss,Fukuyama:2005gg,Malinsky:2005bi}.
They, too, will lead to enhanced LFV rates.  However, both for
definiteness and simplicity, here we focus our discussion on the
case of the simplest \21 inverse seesaw model which we take as a
reference model.
First, we derive a formula for the \m conversion branching ratio in
terms of \lfv parameters of the effective Lagrangian of the model. The
transformation of this Lagrangian, first to the nucleon and then to
the nuclear level, needs special attention to the effects of nucleon
and nuclear structure. The nucleon structure is taken into account on
the basis of the QCD picture of baryon masses and experimental data on
certain hadronic parameters.  The nuclear physics, which is involved
in the muon-nucleus overlap integrals
\cite{faessler:2000pn,Schwieger:1998dd} is evaluated paying special
attention on specific nuclei that are of current experimental interest
like, $^{27}$Al, $^{48}$Ti and $^{197}$Au.


\section{Inverse seesaw mechanism}
\label{sec:inverse-sees-mech}


The model extends minimally the particle content of the Standard
Model by the sequential addition of a pair of two-component \21
singlet leptons, as follows
\begin{equation}
\left(\begin{array}{c}
   \nu_i \\
   e_i \\
\end{array}\right), e_i^c, \nu_i^c, S_i,
\end{equation}
with \(i\) a generation index running over \(1,2,3\). In addition to
the more familiar right-handed neutrinos characteristic of the
standard seesaw model, the inverse seesaw scheme contains an equal
number of gauge singlet neutrinos \(S_i\).  In the original
formulation of the model, these were superstring inspired \e6
singlets, in contrast to the right-handed neutrinos, members of the
spinorial representation. Recently similar constructions have been
considered in the framework of left--right
symmetry~\cite{Akhmedov:1995vm} or SO(10) unified
models~\cite{Malinsky:2005bi,Barr:2005ss,Fukuyama:2005gg}.

In the \(\nu,\nu^c,S\) basis, the \(9\times9\) neutral leptons
mass matrix $\mathcal{M}$ is given as
\begin{equation}
\label{eqn:invSeesaw}
{\mathcal M}=\left(
   \begin{array}{ccc}
      0   & m_D^T & 0   \\
      m_D & 0     & M^T \\
      0   & M     & \mu
   \end{array}\right),
\end{equation}
where \(m_D\) and \(M\) are arbitrary \(3\times3\) complex
matrices in flavour space, whereas \(\mu\) is complex symmetric.
The matrix \(\mathcal{M}\) can be diagonalized by a unitary mixing
matrix \(U_\nu\),
\begin{equation}
   U_\nu^T {\mathcal M} U_\nu = \mathrm{diag}(m_i,M_4,...,M_9),
\end{equation}
yielding 9 mass eigenstates \(n_a\). In the limit of small $\mu$ three
of these correspond to the observed light neutrinos with masses
\(m_i\), while the three pairs of two-component leptons
\((\nu^c_i,S_i)\) combine to form three heavy leptons, of the
quasi-Dirac type~\cite{valle:1983yw}.

The light neutrino mass states \(\nu_i\) are given in terms of the
flavour eigenstates via the unitary matrix $U_\nu$
\begin{equation}
   \nu_i = \sum_{a=1}^{9}(U_\nu)_{ia} n_a.
\end{equation}
which has been studied in earlier
papers~\cite{bernabeu:1987gr,branco:1989bn}. The diagonalization
results in an effective Majorana mass matrix for the light
neutrinos~\cite{gonzalez-garcia:1989rw},
\begin{equation}
\label{eqn:lightNu}
    m_\nu = {m_D^T M^{T}}^{-1} \mu M^{-1} m_D,
\end{equation}
where we are assuming \(\mu, m_D \ll M\).  One sees that the neutrino
masses vanish in the limit $\mu \to 0$ where lepton number
conservation is restored. In models where lepton number is
spontaneously broken by a vacuum expectation value
$\vev{\sigma}$~\cite{gonzalez-garcia:1989rw} one has $\mu = \lambda
\vev{\sigma}$.  Typical parameter values may be estimated from the
required values of the light neutrino masses indicated by oscillation
data~\cite{Maltoni:2004ei} as
\begin{equation}
    \left(\frac{m_\nu}{0.1\mathrm{eV}}\right) =
        \left(\frac{m_D}{100\mathrm{GeV}}\right)^2
        \left(\frac{\mu}{1\mathrm{keV}}\right)
        \left(\frac{M}{10^4\mathrm{GeV}}\right)^{-2}
\label{eqn:lightNuNumeric},
\end{equation}
For typical Yukawas $\lambda \sim 10^{-3}$ one sees that $\mu = 1$ keV
corresponds to a low scale of L violation, $\vev{\sigma} \sim 1$ MeV
(for very low values of $\vev{\sigma}$ this might lead to interesting
signatures in neutrinoless double beta
decays~\cite{berezhiani:1992cd})~\footnote{ Note that such a low scale
  is protected by gauge symmetry.}.

In contrast, in the conventional seesaw mechanism without the
gauge singlet neutrinos \(S_i\) one would have
\begin{equation}
\label{eqn:Seesaw} \left(
   \begin{array}{cc}
      0   & m_D^T   \\
      m_D & M_R
   \end{array}\right),\qquad m_D \ll M_R \Rightarrow  m_\nu = m_D^T M_R^{-1} m_D\,.
\end{equation}
Note that in the ``inverse seesaw'' scheme the three pairs of singlet
neutrinos have masses of the order of \(M\) and their admixture in the
light neutrinos is suppressed as \(\frac{m_D}{M}\). It is crucial to
realize that the mass \(M\) of our heavy leptons can be much smaller
than the \(M_R\) characterizing the right-handed neutrinos in the
conventional seesaw, since the suppression in
Eq.~(\ref{eqn:lightNuNumeric}) is quadratic in $M^{-1}$ (as opposed to
the linear dependence in $M_R^{-1}$ given by Eq.~(\ref{eqn:Seesaw})),
and since we have the independent small parameter \(\mu\)
characterizing the lepton number violation scale.  As a result the
value of \(M\) may be as low as the weak scale (if light enough, these
neutral leptons could give signatures at accelerator experiments
\cite{Dittmar:1990yg,Abreu:1997pa}).

Without loss of generality one can assume \(\mu\) to be diagonal,
\begin{equation}
    \mu = \textrm{diag}\;\mu_i,
\end{equation}
and using the diagonalizing matrix \(U\) of the effective light
neutrino mass matrix \(m_\nu\),
\begin{equation}
    U^T m_\nu U = \textrm{diag}\;m_i,
\end{equation}
equation (\ref{eqn:lightNu}) can be written as
\begin{equation}
    \mathbf{1} = \textrm{diag}\;\sqrt{m_i^{-1}} \cdot U^T m_D^T {M^T}^{-1}
                 \cdot \textrm{diag}\;\sqrt{\mu_i} \cdot
                 \textrm{diag}\;\sqrt{\mu_i} \cdot M^{-1} m_D U
                 \cdot \textrm{diag}\;\sqrt{m_i^{-1}}.
\end{equation}
In the basis where the charged lepton Yukawa couplings are
diagonal the lepton mixing matrix is simply the rectangular matrix
formed by the first three rows of
\(U_\nu\)~\cite{schechter:1980gr}.

In analogy to the standard seesaw mechanism \cite{Casas:2001sr} it
is thus possible to define a complex orthogonal matrix
\begin{equation}
\label{eqn:R}
    R = \textrm{diag}\;\sqrt{\mu_i} \cdot M^{-1} m_D U \cdot \textrm{diag}\;\sqrt{m_i^{-1}}
\end{equation}
with 6 real parameters.  Using \(R\), the neutrino Yukawa coupling
matrix \(Y_\nu = \frac{1}{v\sin\beta} m_D\) can be expressed as
\begin{equation}
\label{eqn:R}
    Y_\nu = \frac{1}{v\sin\beta} M \cdot \textrm{diag}\;\sqrt{\mu_i^{-1}} \cdot R
            \cdot \textrm{diag}\;\sqrt{m_i} \cdot U^\dagger,
\end{equation}
To further simplify our discussion we make the assumption that the
eigenvalues of both \(M\) and \(\mu\) are degenerate and that
\(R\) is real. This allows us to easily compare our results with
those obtained previously in
Ref.~\cite{Deppisch:2002vz,Deppisch:2003wt} for the case of the
conventional seesaw mechanism.


\section{The effective quark-level Lagrangian}

\begin{figure}[t]
\centering
\includegraphics[clip,width=0.35\linewidth]{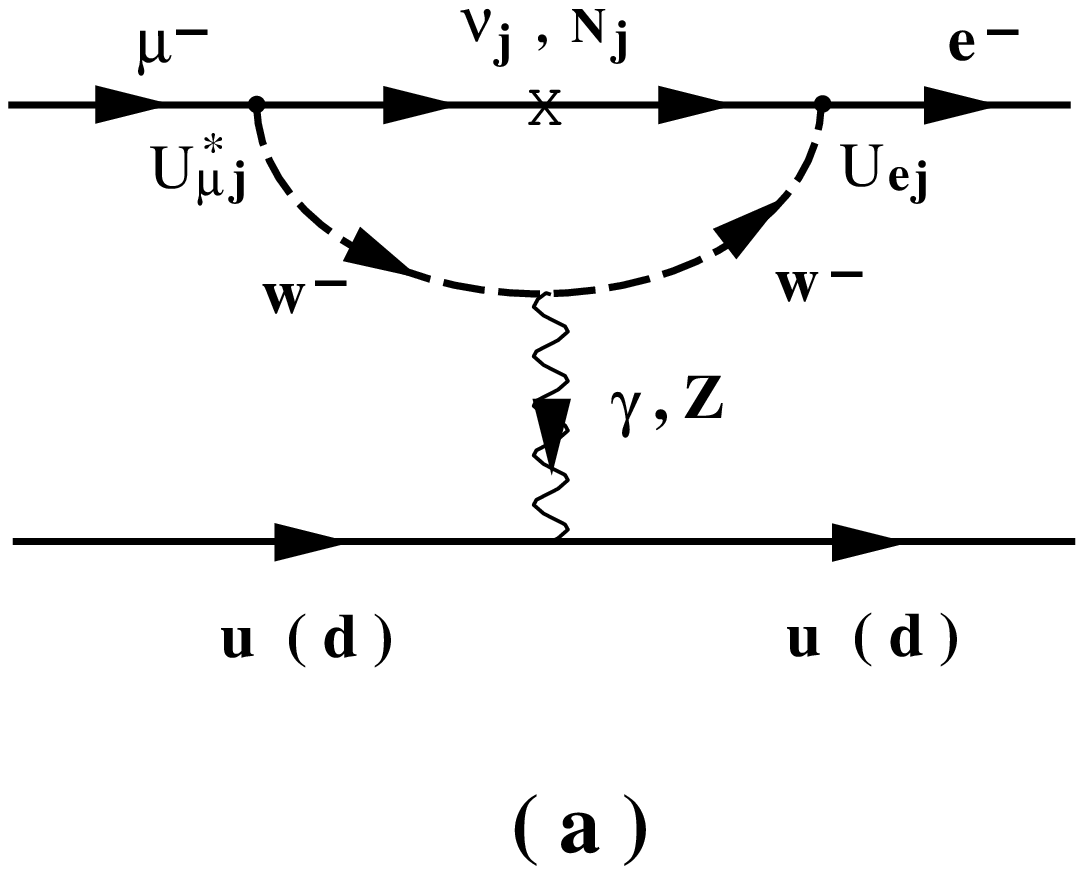}
\includegraphics[clip,width=0.35\linewidth]{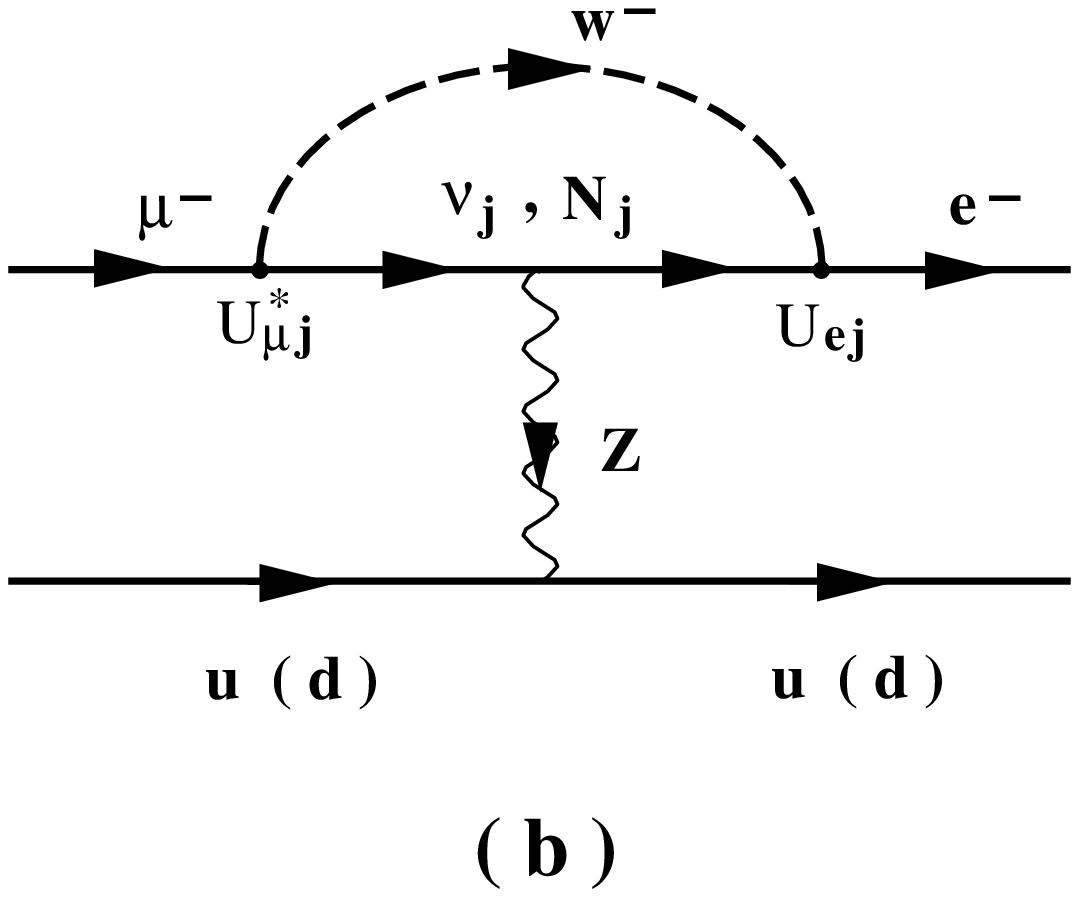}\\
\includegraphics[clip,width=0.7\linewidth]{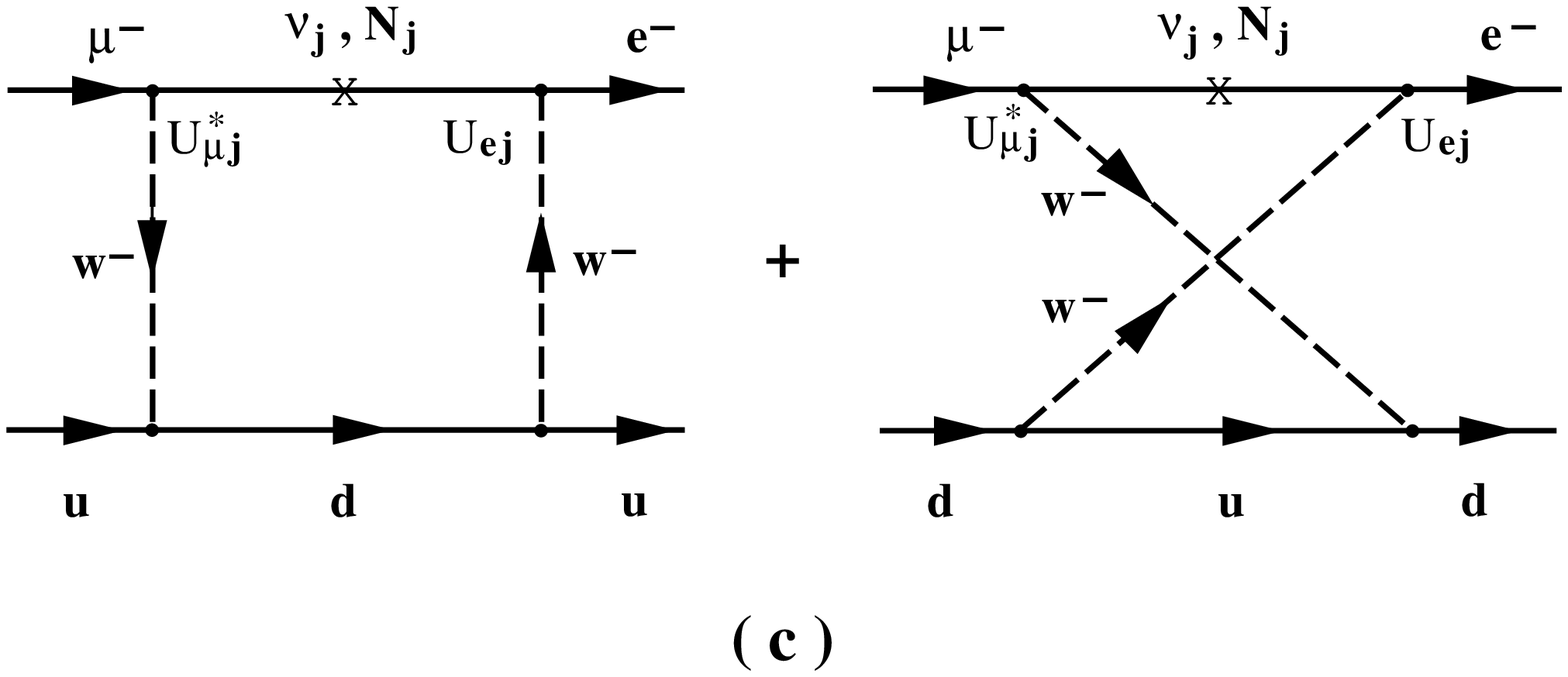}
\caption{Diagrams for $\mu-e$ conversion: photonic (a), Z-boson (a,b)
  and W-boson (c) exchange.}
    \label{fig:Br_M_mu}
\end{figure}

In our model the \lfv arises from penguin photon and Z exchange as
well as box diagrams, as illustrated in Fig.~\ref{fig:Br_M_mu}.
The resulting effective Lagrangian can be expressed
as~\cite{Hisano:1995cp}
\begin{eqnarray}
   {\mathcal{L}}_{eff}&=&{\mathcal{L}}_{eff}^\gamma +
   {\mathcal{L}}_{eff}^Z + {\mathcal{L}}_{eff}^{box}\\
   {\mathcal{L}}_{eff}^\gamma &=&
        -\frac{e^2}{q^2}\bar e
        \left[q^2\gamma_\alpha(A^L_1 P_L + A^R_1 P_R)
            + m_\mu i\sigma_{\alpha\beta}q^\beta(A^L_2 P_L + A^R_2 P_R)
        \right]\mu \times\sum_q Q^q\bar q \gamma^\alpha q \label{eqn:EffectiveLagrangianPhoton}\\
        &=&-\frac{e^2}{q^2}\sum_q\left[\frac{1}{2}(A^L_1 +
        A^R_1)j_\alpha^V+\frac{1}{2}(A^R_1
        -
        A^L_1)j_\alpha^A
        + \frac{i}{2}(A^L_2 + A^R_2)m_\mu j_{\alpha\beta}q^\alpha\right]Q^qJ^{V\beta}_{(q)}\label{eqn:EffectiveLagrangianPhoton2}\\
   {\mathcal{L}}_{eff}^Z &=&
        \frac{g_Z^2}{m_Z^2}\bar e
        \left[\gamma_\alpha(F^L P_L + F^R P_R)\right]\mu
        \times\sum_q \frac{Z^q_L+Z^q_R}{2}\bar q \gamma^\alpha q \label{eqn:EffectiveLagrangianZ}\\
        &=&\frac{g_Z^2}{m_Z^2}\sum_q\left[\frac{1}{2}(F^L + F^R)j_\alpha^V+\frac{1}{2}(F^R
        -
        F^L)j_\alpha^A\right]\frac{Z^q_L+Z^q_R}{2}J^{V\alpha}_{(q)}\label{eqn:EffectiveLagrangianZ2}\\
\label{eqn:EffectiveLagrangianBox}
  {\mathcal{L}}_{eff}^{box} &=&
        e^2
        \bar e\left[\gamma_\alpha(D^L_q P_L + D^R_q P_R)\right]\mu
        \times\sum_q Q^q\bar q \gamma^\alpha q \\
        &=&e^2\sum_q\left[\frac{1}{2}(D^L_q + D^R_q)j_\alpha^V+\frac{1}{2}(D^R_q
        - D^L_q)j_\alpha^A\right]Q^q J^{V\alpha}_{(q)}\label{eqn:EffectiveLagrangianBox2}
\end{eqnarray}
with \(Q^q\) the electric charge of quark \(q\), and
\begin{equation}
    Z^q_{L/R} = (T^q_3)_{L/R} - Q^q\sin^2\theta_W.
\end{equation}
The expressions in Eqs.~(\ref{eqn:EffectiveLagrangianPhoton2}),
(\ref{eqn:EffectiveLagrangianZ2}) and
(\ref{eqn:EffectiveLagrangianBox2}) correspond to the notation in
Equation (5) of Ref.~\cite{Kosmas:2001mv}.  The coefficients
\(A_1^{L/R}\), \(A_2^{L/R}\), \(F^{L/R}\), \(D_q^{L/R}\) which
give rise to lepton flavour violation, are given by (for \(\mu-e\)
conversion, the indices \(i,j\) are always \(i=2 (\sim \mu)\) and
\(j=1 (\sim e)\), and are thus omitted for simplicity in the above
formulae) \cite{Ilakovac:1994kj}:
\begin{eqnarray}
    A_1^{L}  &=& \sum_{i=1}^{9}U_{2i}^*U_{1i}F_\gamma(\lambda_i),\\
    A_1^R    &=& 0\\
    A_2^{L}  &=&
        \frac{m_e}{m_\mu}\sum_{i=1}^{9}U_{2i}^*U_{1i}G_\gamma(\lambda_i),\\
    A_2^{R}  &=& \sum_{i=1}^{9}U_{2i}^*U_{1i}G_\gamma(\lambda_i),\\
    F^{L} &=&
    \sum_{i,j=1}^{9}U_{2i}^*U_{1i}\left(F_Z(\lambda_i)
        +C_{ij}H_Z(\lambda_i,\lambda_j)+C_{ij^*}G_Z(\lambda_i,\lambda_j)\right),\\
    F^{R} &=& 0,\\
    D^{L}_q &=&
        \sum_{i,j=1}^{9}\left(U_{2i}^*U_{1i}F_{box}(\lambda_i,\lambda_j)
        + U_{2i}U_{1i}^*G_{box}(\lambda_i,\lambda_j)\right),\\
    D^{R}_q &=& 0,
\end{eqnarray}
with
\begin{equation}\label{eqn:CMixingMatrix}
    C_{ij}=\sum_{k=1}^3U_{ki}U^*_{kj}.
\end{equation}
The corresponding form-factor functions in the above terms are
given by \cite{Ilakovac:1994kj}
\begin{eqnarray}
    F_\gamma(x) &=& \frac{7x^3-x^2-12x}{12(1-x)^3}-\frac{x^4-10x^3+12x^2}{6(1-x)^4}\ln x,\\
    G_\gamma(x) &=& -\frac{2x^3+5x^2-x}{4(1-x)^3}-\frac{3x^3}{2(1-x)^4}\ln x,\\
    F_Z(x)      &=& -\frac{5x}{2(1-x)}-\frac{5x^2}{2(1-x)^2}\ln x,\\
    G_Z(x,y)    &=& -\frac{1}{2(x-y)}\left[\frac{x^2(1-y)}{1-x}\ln x - \frac{y^2(1-x)}{1-y}\ln y\right],\\
    H_Z(x,y)    &=& \frac{\sqrt{xy}}{4(x-y)}\left[\frac{x^2-4x}{1-x}\ln x - \frac{y^2-4y}{1-y}\ln y\right],\\
    F_{box}(x,y)&=& \frac{1}{x-y}\left[\left(1+\frac{xy}{4}\right)\left(\frac{1}{1-x}
         + \frac{x^2\ln x}{(1-x)^2} - \frac{1}{1-y} -\frac{y^2\ln
         y}{(1-y)^2}\right)\right. \nonumber\\
                &-& 2xy\left.\left(\frac{1}{1-x} +\frac{x\ln x}{(1-x)^2} - \frac{1}{1-y} -\frac{y\ln
         y}{(1-y)^2}\right)\right],\\
    F_{box}(x,y)&=& -\frac{\sqrt{xy}}{x-y}\left[\left(4+xy\right)\left(\frac{1}{1-x} +
          \frac{x\ln x}{(1-x)^2} - \frac{1}{1-y} -\frac{y\ln
         y}{(1-y)^2}\right)\right. \nonumber\\
                &-& 2\left.\left(\frac{1}{1-x} +\frac{x^2\ln x}{(1-x)^2} - \frac{1}{1-y} -\frac{y^2\ln
         y}{(1-y)^2}\right)\right].
\end{eqnarray}

The effective Lagrangians for the $\mu-e$ diagrams of Eqs.
(\ref{eqn:EffectiveLagrangianPhoton}),
(\ref{eqn:EffectiveLagrangianZ}) and
(\ref{eqn:EffectiveLagrangianBox}) can be compactly written as
\begin{eqnarray}
{\cal L}_{eff}^{q}\ =\  G_a \left(\sum_{A,B;q}\ \eta_{AB}^{(q)}
j_{\mu}^A\ J_{(q)}^{B\mu} + \sum_{C,D;q} \eta_{CD}^{(q)} j^C\
J_{(q)}^{D} + \sum_{q} \eta_{T}^{(q)} j_{\mu\nu}\
J_{(q)}^{\mu\nu}\right) \:, \quad a=ph,nph,\label{eff-q}
\end{eqnarray}
where summation involves $A,B = \{A,V\}$; $C,D =\{S,P\}$ and $q=
\{u,d,s\}$. The coupling strength factor \(G_a\) is given by
\(G_{nph}=\frac{G_F}{\sqrt{2}}\) in the non-photonic and by
\(G_{ph}=\frac{4\pi\alpha}{q^2}\) in the photonic case. The
parameters $\eta_i^{(q)}$ depend on the specific \lfv model
assumed. The lepton and quark currents are
\begin{eqnarray}
j_{\mu}^V = \bar e \gamma_{\mu} \mu,
j_{\mu}^A = \bar e \gamma_{\mu}\gamma_5 \mu,  \\
j^S = \bar e \ \mu,
j^P = \bar e \gamma_{5} \mu, \\
j_{\mu\nu} = \bar e \sigma_{\mu\nu} \mu,\\
J_{(q)}^{V\mu} = \bar q \gamma^{\mu} q,
J_{(q)}^{A\mu} = \bar q \gamma^{\mu}\gamma_5 q, \\
J_{(q)}^{S} = \bar q \ q,
J_{(q)}^{P} = \bar q \gamma_5\ q, \\
J_{(q)}^{\mu\nu} = \bar q \sigma^{\mu\nu} q.
\label{currents-q}
\end{eqnarray}
In our model, the only nonvanishing contributions are
$\eta_{VV}^{(q)}$ and $\eta_{AV}^{(q)}$,
\begin{eqnarray}
\eta_{VV}^{(q)}&=& \frac{1}{2}(F^L+F^R)(Z_L^q+Z_R^q)+\frac{1}{2}Q^q(D_q^L+D_q^R),\\
\eta_{AV}^{(q)}&=&
\frac{1}{2}(F^L-F^R)(Z_L^q+Z_R^q)+\frac{1}{2}Q^q(D_q^L-D_q^L).
\end{eqnarray}
in the non-photonic case and
\begin{eqnarray}
\eta_{VV}^{(q)}&=& \frac{1}{2}Q^q(A_1^L+A_1^R),\\
\eta_{AV}^{(q)}&=& \frac{1}{2}Q^q(A_1^L-A_1^R),
\end{eqnarray}
in the photonic case.

\section{The effective nucleon-level Lagrangian}

The nucleon level effective Lagrangian obtained through the
reformulation of the quark level effective Lagrangian (\ref{eff-q})
can be written in terms of the effective nucleon fields and the
nucleon isospin operators as
\begin{eqnarray}
\label{eff-N}
{\cal L}_{eff}^{N}\ &=&\   G_a
\sum_{A,B}\  j_{\mu}^A \left(\alpha_{AB}^{(0)} J_{(0)}^{B\mu} +
\alpha_{AB}^{(3)} J_{(3)}^{B\mu}\right).
j^C (\alpha_{CD}^{(0)} J_{(0)}^{D} +
\alpha_{CD}^{(3)} J_{(3)}^{D}) +\\
\label{eff-N2}
&+& \left. j_{\mu\nu} (\alpha_{T}^{(0)} J_{(0)}^{\mu\nu} +
\alpha_{T}^{(3)} J_{(3)}^{\mu\nu})\right], \qquad a=ph,nph.
\end{eqnarray}
The isoscalar  $J_{(0)}$ and isovector $J_{(3)}$ nucleon currents are defined as
\begin{eqnarray}\nn
J_{(k)}^{V\mu} = \bar N \gamma^{\mu}\tau_k N, \ \
J_{(k)}^{A\mu} = \bar N \gamma^{\mu}\gamma_5 \tau_k N,\ \
J_{(k)}^{S} = \bar N \tau_k N, \ \
J_{(k)}^{P} = \bar N \gamma_5 \tau_k N,\ \
J_{(k)}^{\mu\nu} = \bar N \sigma^{\mu\nu}\tau_k N,
\label{Nucl-curr}
\end{eqnarray}
where $ k = 0,3$ and $\tau_0 \equiv \hat I$.

The relationship between the coefficients $\alpha$ in
Eq.~(\ref{eff-N}) and the fundamental \lfv parameters $\eta_{AB}$
of the quark level Lagrangian (\ref{eff-q}) can be found as
follows. We start from the equations which relate the various
nucleon form factors $G_{K}^{(q,N)}$ with matrix elements of the
quark states and those of the nucleon states
\begin{eqnarray}\label{mat-el1}
\langle N|\bar{q}\ \Gamma_{K}\ q|N\rangle = G_{K}^{(q,N)}
\bar{\Psi}_N\ \Gamma_{K}\ \Psi_N,
\end{eqnarray}
with $q=\{u,d,s\}$,  $N=\{p,n\}$ and  $K = \{V,A,S,P\}$,
$\Gamma_K = \{\gamma_{\mu}, \gamma_{\mu}\gamma_5, 1, \gamma_5\}$.
Since the maximum momentum transfer in $\mu -e$ conversion is much
smaller than the typical scale, we may neglect the ${\bf
q}^{2}$-dependence of $G_{K}^{(q,N)}$. Assuming isospin symmetry,
we find
\begin{eqnarray}\label{isosym}
G_{K}^{(u,n)}=G_{K}^{(d,p)}\equiv G_{K}^{d}, \ \ \
G_{K}^{(d,n)}=G_{K}^{(u,p)}\equiv G_{K}^{u},\ \ \
G_{K}^{(s,n)}=G_{K}^{(s,p)}\equiv G_{K}^{s}.
\end{eqnarray}

For the coherent nuclear $\mu^--e^-$ conversion, only the vector
and scalar nucleon form factors are needed (the axial and
pseudoscalar nucleon currents couple to the nuclear spin and for
spin zero nuclei they can contribute only to the incoherent
transitions). The vector current form factors are determined
through the assumption of conservation of vector current at the
quark level which gives
\begin{eqnarray}\label{cvc}
G_{V}^{u}=2,\ \ \  G_{V}^{d}=1, \ \ \ G_{V}^{s}=0.
\end{eqnarray}

The coefficients $\alpha^{(\tau)}_{AB}$ of the nucleon level
Lagrangian (\ref{eff-N}) can be expressed in terms of the \lfv
parameters $\eta_{AB}$ of the quark level effective Lagrangian in
Eq. (\ref{eff-q}) as
\begin{eqnarray}
\label{alpha} \alpha_{IV}^{(0)} &=&\frac{1}{2}(\eta_{IV}^{(u)} +
\eta_{IV}^{(d)}) (G_{V}^{u} + G_{V}^{d}),\nn\\
\alpha_{IV}^{(3)} &=& \frac{1}{2}(\eta_{IV}^{(u)} -
\eta_{IV}^{(d)}) (G_{V}^{u} - G_{V}^{d}),
\end{eqnarray}
where $I=V,A$.

From the Lagrangian (\ref{eff-N}), following standard procedure,
we can derive a formula for the total $\mu-e$ conversion branching
ratio. In this paper we restrict ourselves to the coherent process
which is the dominant channel of $\mu-e$ conversion. For most
experimentally interesting nuclei, this accounts for more than
$90\%$ of the total \m{} branching ratio \cite{Schwieger:1998dd}.
To leading order of the non-relativistic reduction the coherent
$\mu-e$ conversion branching ratio takes the form
\begin{equation}
R_{\mu e^-}^{\mathrm{coh}} \ = \ {\cal Q}_a G_a^2 \frac{p_e E_e}
{2 \pi } \ \frac{ {\cal M}_a^2 } { \Gamma ({\mu^-\to
\mathrm{capture}}) } \, \qquad \mathrm{a=ph,nph} \, , \label{Rme}
\end{equation}
where $p_e, E_e$ are the outgoing electron 3-momentum and energy
and ${{\cal M}}^2_{ph}$ (${{\cal M}}^2_{nph}$) represent the
squares of the nuclear matrix elements for the photonic and
non-photonic modes of the process. The quantity \({\mathcal Q}_a\)
is defined as
\begin{eqnarray}
{\cal Q}_a = |\alpha_{VV}^{(0)}+\alpha_{VV}^{(3)}\ \phi|^2 +
|\alpha_{AV}^{(0)}+\alpha_{AV}^{(3)}\ \phi|^2, \label{Rme.1}
\end{eqnarray}
with the corresponding coefficients \(\alpha\) for the the
photonic and non-photonic contributions and depends on the nuclear
parameters through the factor
\begin{eqnarray}\label{phi}
\phi = ({\cal M}_p - {\cal M}_n)/({\cal M}_p + {\cal M}_n) \, .
 \end{eqnarray}
The ${\cal M}_{p,n}$ are given by
\begin{equation}
{\cal M}_{p,n} = 4\pi \int (g_e g_\mu + f_e f_\mu ) \rho_{p,n} (r)
r^2  dr. \label{V.1}
\end{equation}
In the latter equation, $\rho_{p,n}(r)$ are the spherically
symmetric proton (p) and neutron (n) nuclear densities normalized
to the atomic number $Z$ and neutron number $N$, respectively, of
the target nucleus.  Here $g_\mu$, $f_\mu$ are the top and bottom
components of the $1s$ muon wave function and $g_e$, $f_e$ are the
corresponding components of the Coulomb modified electron wave
function \cite{pana-kos,Kitano:2002mt}.

In the present work, the matrix elements ${\cal M}_{p,n}$, defined in
Eq. (\ref{V.1}), have been numerically calculated using proton
densities $\rho_{p}$ from Ref.  \cite{DeJager:1987qc} and neutron
densities $\rho_{n}$ from Ref.  \cite{Gibbs:1987fd}. The muon wave
functions $f_\mu$ and $g_\mu$ (and also $g_e$, $f_e$) were obtained by
solving the Dirac equation with the Coulomb potential produced by the
densities $\rho_{p,n}$ by using artificial neural network techniques.
In this way, relativistic effects and vacuum polarization corrections
have been taken into account \cite{pana-kos,Kitano:2002mt}. The latter
method has recently been applied for evaluating the \(\tau^-\) wave
functions in a set of (medium-heavy and heavy) nuclei for obtaining
the \(\tau\)-capture rate by nuclei.

\section{Results and discussion}

The results for ${\cal M}_{p,n}$ corresponding to a set of nuclei
throughout the periodic Table including systems with good
sensitivity to the $\mu -e$ conversion are shown in
table~\ref{tab:results}. In this table we also present the muon
binding energy $\epsilon_b$ and the experimental values for the
total rate of the ordinary muon capture $\Gamma_{\mu c}$
\cite{Suzuki:1987jf}. We give the ingredients required to
determine the branching ratio $R_{\mu e}$ for the three nuclei Al,
Ti and Au, of current experimental interest.

As has been discussed in Ref. \cite{Kosmas:1990tc}, for the
description of the long range {photonic} contribution only the
proton matrix elements ${\cal M}_{p}$ are required. In the case of
the non-photonic mechanisms (short-range contributions), however,
both protons and neutrons contribute and therefore both ${\cal
M}_{p,n}$ matrix elements are needed. The latter are obtained by
using densities extracted from the data on pionic atoms or the
pion-nucleus scattering \cite{Gibbs:1987fd}.
\begin{figure}[t]
\centering
\includegraphics[clip,width=0.49\linewidth]{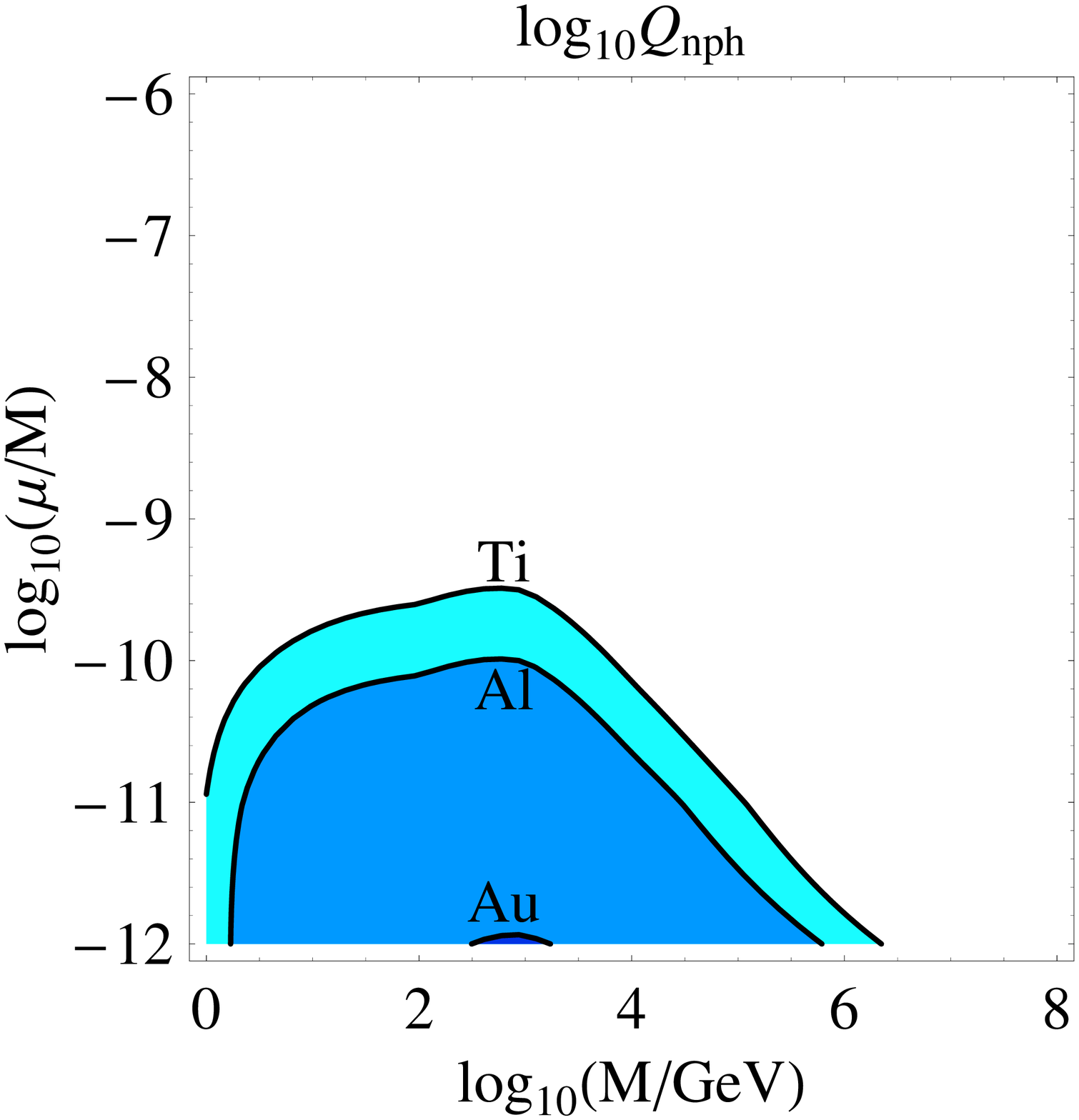}
\includegraphics[clip,width=0.49\linewidth]{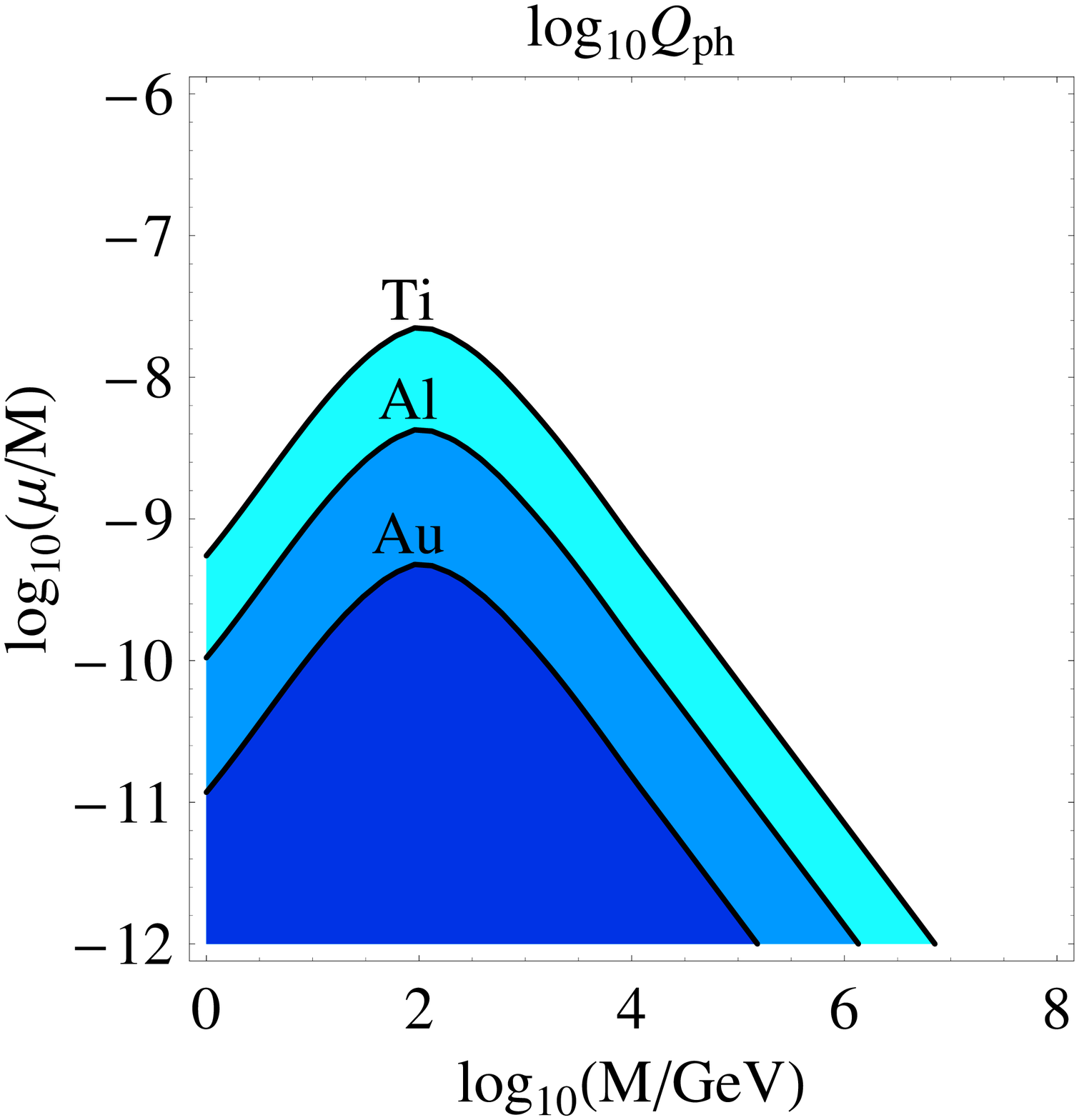}
\caption{Present and expected limits on the model parameters \(M\) and
  \(\mu/M\).  The shaded areas are excluded by the bounds on
  \(Q_\text{nph}\) (left panel) and \(Q_\text{ph}\) (right panel)
  given in Table~\ref{tab:Q_bounds}.}
     \label{fig:Q_nph}
\end{figure}
\begin{figure}[h]
  \centering
\includegraphics[clip,height=6cm,width=0.49\linewidth]{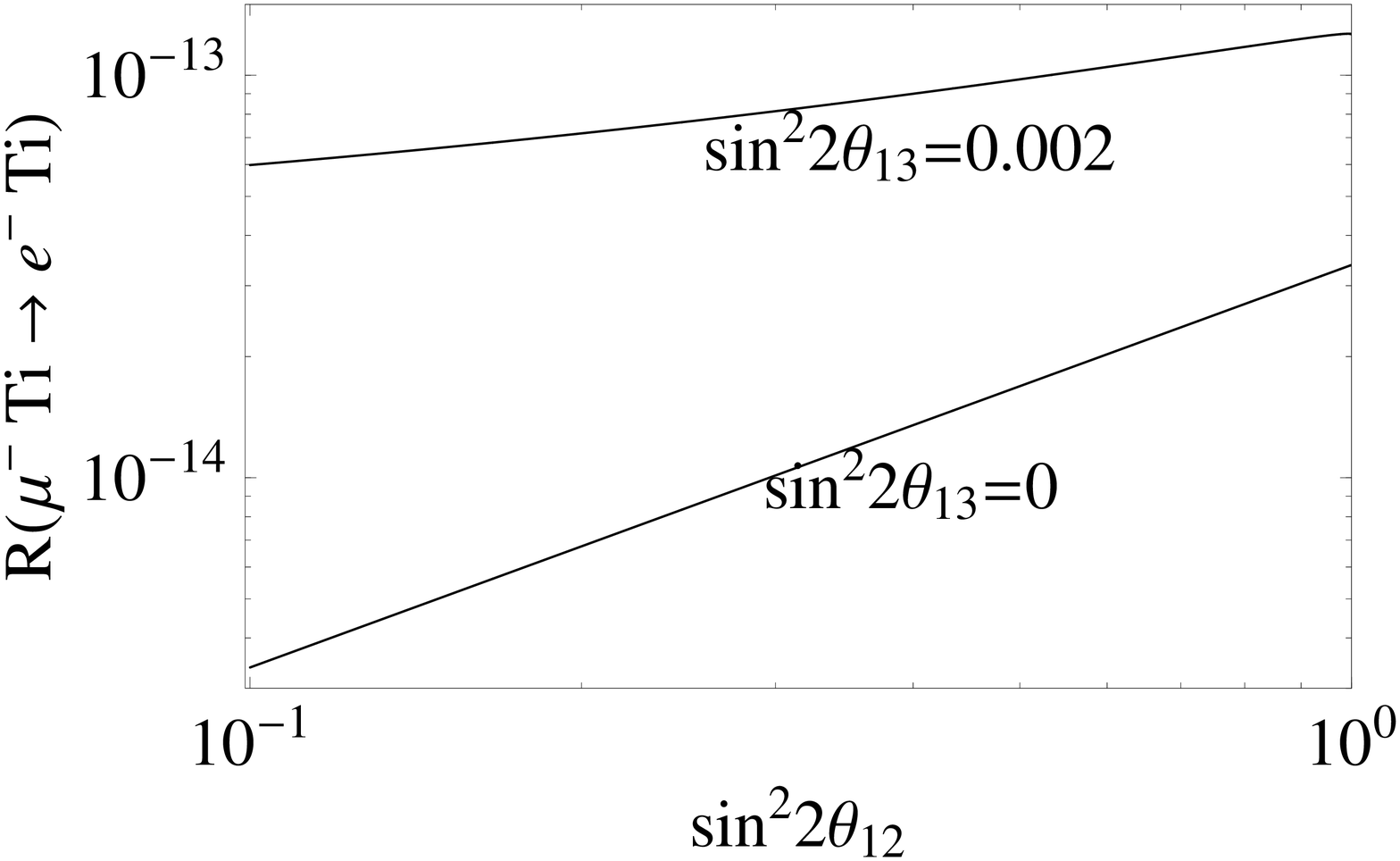}
\includegraphics[clip,height=6cm,width=0.49\linewidth]{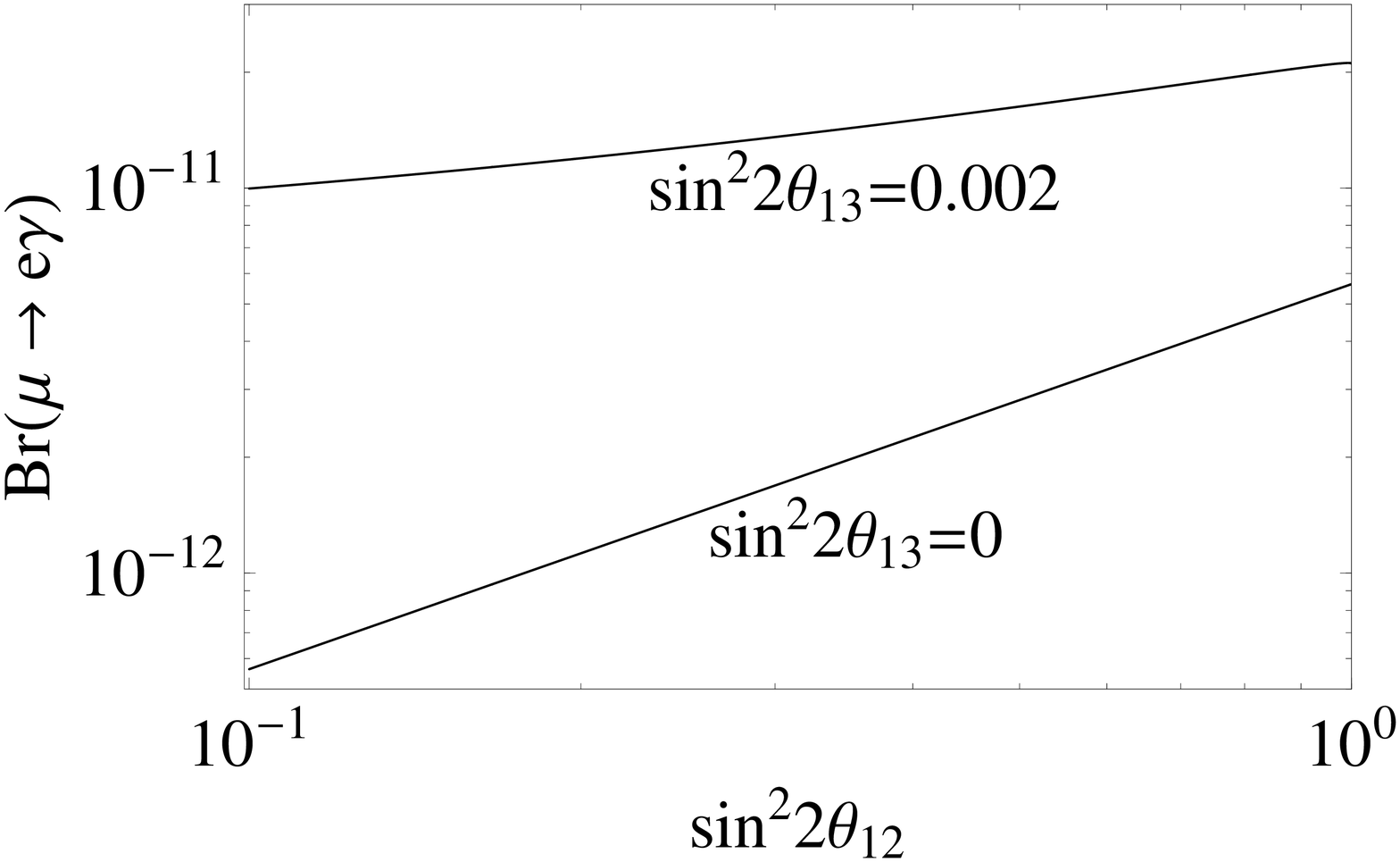}
\caption{Relation of branching ratios for $\mu^--e^-$ conversion (left
  panel) and $\mu^-\to e^-\gamma$ (right panel) with the solar
  neutrino mixing angle, for different values of $\theta_{13}$. The
  inverse seesaw parameters are given by: \(M=100\)~GeV,
  \(\mu=10\)~eV.}
\label{fig:rel-sol}
\end{figure}
\begin{figure}[h]
  \centering
  \includegraphics[clip,height=6cm,width=0.49\linewidth]{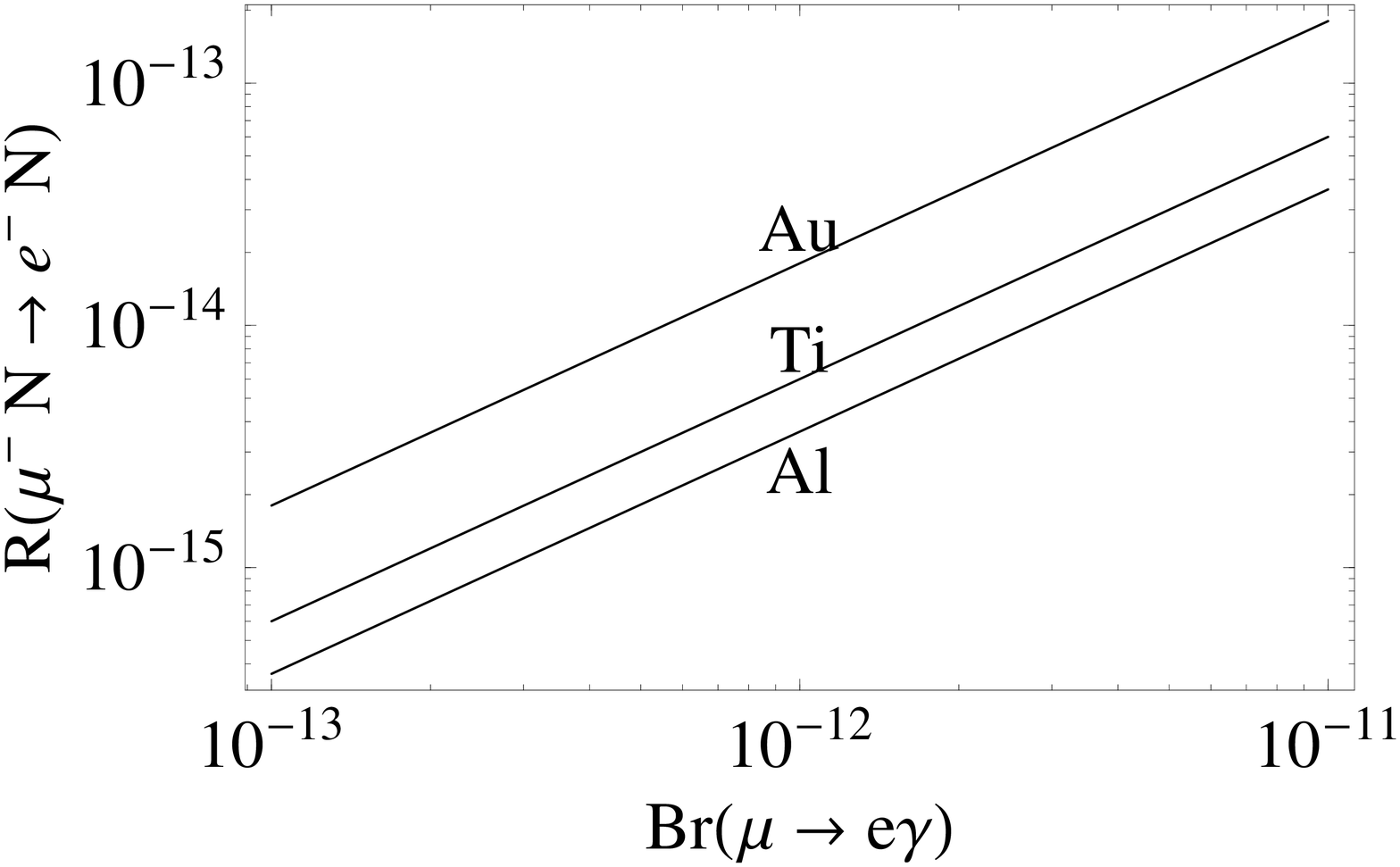}
  \caption{Correlation between nuclear $\mu^--e^-$ conversion and
    $\mu^-\to e^-\gamma$ decay in the inverse seesaw model. }
\label{fig:corr}
\end{figure}

Using the values of Table 1 and the existing
\cite{vanderSchaaf:2003ti} or expected
\cite{Molzon:1998kg,Kuno:2000kd} experimental sensitivities on
$R_{\mu
  e}$, we can derive the corresponding sensitivity upper limits on the
particle physics parameters characterizing the effective Lagrangians
(\ref{eff-q}) and (\ref{eff-N}).
The most straightforward limits can be set on the quantities
${\cal Q}_a$ of Eq. (\ref{Rme})
which are given in Table \ref{tab:Q_bounds}.

In order to achieve limits on the particle physics leading to
\(\mu-e\) conversion, the quark level effective Lagrangian of the
model is adjusted to the form of Eq. (\ref{eff-q}) and by
identifying the effective parameters $\eta_{AB}^{(q)}$ with
expressions in terms of model parameters. This way the upper
bounds on $Q_a$ from Table \ref{tab:Q_bounds} can be translated to
restrictions on the model parameters present in these expressions.
Figure~\ref{fig:Q_nph} shows the sensitivity bounds on the
parameters \(M\) and \(\mu\) characterizing the inverse seesaw
model, that follow from experiments with Al, Au and Ti targets,
respectively.

It is also interesting to explore how these rates depend on the
relevant neutrino mixing angles which are probed in solar neutrino
experiments. This is shown in figure~\ref{fig:rel-sol}, where the
relation between \(R(\mu^-\text{Ti}\to e^-\text{Ti})\)
(\(Br(\mu\to e\gamma)\)) and the solar neutrino angle
\(\theta_{12}\) for fixed \(M\) and \(\mu\) is displayed. As can
be seen in the figure, there is also a strong dependence on the
small neutrino mixing angle \(\theta_{13}\).

In designing future experiments testing for \lfv it is instructive
to determine how the branching ratios for $\mu^--e^-$ conversion
and the $\mu^-\to e^-\gamma$ are related. In Fig.~\ref{fig:corr}
we show explicitly that, in the inverse seesaw model, the rates
for $\mu^--e^-$ conversion and that for the $\mu^-\to e^-\gamma$
decay are strongly correlated, indicating the dominance of the
photonic diagram in Fig.~\ref{fig:Diagrams}(a).

\section{Summary  and Conclusions}
In the present paper we constructed an effective Lagrangian
describing the photonic and non-photonic $\mu^--e^-$ conversion in
the context of the inverse seesaw model and specified the \lfv
parameters characterizing this process.
We focused in the simplest inverse seesaw model.  The interest in
the model is that it accounts for the observed masses and mixings
indicated by current oscillation data in such a way that the
underlying physics ``does not decouple'' and can be
phenomenologically probed experimentally.  The model provides a
framework for enhanced \lfv rates with a rather simple, almost
minimalistic, particle content.  In contrast to the conventional
seesaw, this is achieved without need of supersymmetrization.
We derived a general formula for the coherent $\mu^--e^-$ conversion
branching ratio in terms of the \lfv parameters of the above quark
level effective Lagrangian and we calculated the corresponding nuclear
matrix elements of currently interesting nuclear targets like
$^{197}$Au (the current SINDRUM II target), $^{27}$Al (the target of
the ongoing MECO experiment) and $^{48}$Ti (the target for the
upcoming PRISM experiment).
These results are given in table~\ref{tab:Q_bounds} and can be
used to obtain sensitivity limits from existing or planned
experiments on \lfv parameters in a variety of particle physics
models.
We have considered in detail the new important contributions to \m
conversion present in the inverse seesaw model that come from the
exchange of the relatively light \21 singlet neutral leptons.
Figure~\ref{fig:Q_nph} shows the sensitivity bounds on the
parameters \(M\) and \(\mu\) characterizing the inverse seesaw
model, that follow from experiments with Al, Au and Ti,
respectively. On the other hand figure~\ref{fig:rel-sol} displays
the relation of the \lfv rates with the relevant neutrino mixing
angles, while Fig.~\ref{fig:corr} establishes that, in the inverse
seesaw model, the rates for $\mu^--e^-$ conversion and that for
the $\mu^-\to e^-\gamma$ decay are highly correlated.

\bigskip

\bigskip

This work was supported by Spanish grant BFM2002-00345, by the
European Commission Human Potential Program RTN network
MRTN-CT-2004-503369.  T.S.K. and F.D. would like to express their
appreciation to IFIC for hospitality.

\newpage


\newpage
\begin{table}
\begin{center}
\begin{tabular}{|r|c|c|c|c|c|c|}
\hline Nucleus & $|{\bf p}_e| \, (fm^{-1})$ & $\Gamma_{\mu c} \, (
10^{6} \, s^{-1})$ &
${\cal M}^2_{ph} \, (fm^{-3})$ & ${\cal M}^2_{nph} \, (fm^{-3})$  & ${\quad\qquad\cal\phi\qquad\quad}$ \\
\hline
$^{12}$C  & 0.533 &  0.0388 & 0.00007 & 0.00029 &  0.0000  \\
$^{27}$Al & 0.532 &  0.705  & 0.00204 & 0.00821 & -0.0022  \\
$^{32}$S  & 0.531 &  1.352  & 0.00433 & 0.01656 &  0.0225  \\
$^{40}$Ca & 0.529 &  2.557  & 0.00982 & 0.03667 &  0.0350  \\
$^{48}$Ti & 0.528 &  2.590  & 0.01217 & 0.05560 & -0.0645  \\
$^{63}$Cu & 0.524 &  5.676  & 0.02883 & 0.12631 & -0.0445  \\
$^{90}$Zr & 0.517 &  8.660  & 0.06713 & 0.29713 & -0.0493  \\
$^{112}$Cd& 0.511 & 10.610  & 0.08416 & 0.37712 & -0.0552  \\
$^{197}$Au& 0.485 & 13.070  & 0.15571 & 0.68691 & -0.0478  \\
$^{208}$Pb& 0.482 & 13.450  & 0.18012 & 0.80892 & -0.0563  \\
$^{238}$U & 0.474 & 13.100  & 0.19360 & 0.87797 & -0.0608  \\
\hline
\end{tabular} \\[.3cm]
\caption{Ingredients entering Eq. (\ref{V.1}) which gives the
branching ratio of the charged lepton flavour violating $\mu-e$
conversion for a set of nuclei throughout the periodic table. We
note that by neglecting the electron mass we have $E_e \approx p_e
c$.} \label{tab:results}
\end{center}
\end{table}

\vskip -3cm

\begin{table}
\begin{center}
\begin{tabular}{|c|c|c|c|}
\hline
Parameter        & Present limits (PSI)   & Expected limits (MECO) & Expected limits (PRISM)\\
                 & $^{197}$Au             & $^{27}$Al              & $^{48}$Ti              \\
\hline ${\cal Q}_{ph}$  & \(1.96\cdot 10^{-16}\) & \(2.68\cdot
10^{-18}\) & \(8.39\cdot 10^{-20}\)\\
${\cal Q}_{nph}$ & \(4.45\cdot 10^{-15}\) & \(6.67\cdot 10^{-19}\) & \(1.84\cdot 10^{-20}\) \\
\hline
\end{tabular} \\[.3cm]
\caption{Upper bounds on the parameters \(Q_a\) (see text, for
definition)
  inferred from the SINDRUM II data on the $\mu^--e^-$ conversion in
  ${}^{197}$Au [Eq. (\ref{eq:SINDRUM})] as well as from the expected
  sensitivities of the current MECO (BNL) [Eq. (\ref{eq:MECO})] and
  PRISM (KEK) experiments [Eq.~(\ref{eq:PRISM})] with $^{27}$Al and
  $^{48}$Ti stopping targets respectively. } \label{tab:Q_bounds}
\end{center}
\end{table}
%

\end{document}